\documentstyle[11pt,aasms4,rotating]{article}

\def\simlt{\lower.5ex\hbox{$\; \buildrel < \over \sim \;$}}
\def\simgt{\lower.5ex\hbox{$\; \buildrel > \over \sim \;$}}

\def\gcm3{{\rm\,g\,cm^{-3}}}
\def\ncm3{{\rm\,cm^{-3}}}

\def\>{$>$}
\def\<{$<$}

\lefthead{Melia et al.}
\righthead{}

\begin{document}
\centerline{Submitted to the Editor of the Astrophysical Journal Letters}
\vskip 0.5in
\title{\bf An Accretion-Induced X-ray Flare in Sgr A*}

\author{Siming Liu\altaffilmark{1} and Fulvio Melia$^{1,2,3}$}

\affil{$^1$Physics Department, The University of Arizona, Tucson, AZ 85721}
\affil{$^2$Steward Observatory, The University of Arizona, Tucson, AZ 85721}


\altaffiltext{3}{Sir Thomas Lyle Fellow and Miegunyah Fellow.}


\begin{abstract}

The recent detection of a three-hour X-ray flare from Sgr A* by {\it Chandra}
provides very strong evidence for a compact emitting region near this
supermassive black hole at the Galactic center. Sgr A*'s mm/sub-mm spectrum 
and linear polarimetric properties, and its quiescent-state X-ray flux density,
are consistent with a model in which low angular momentum gas captured at
large radii circularizes to form a hot, magnetized Keplerian flow within
tens of Schwarzschild radii of the black hole's event horizon.  In Sgr A*'s
quiescent state, the X-ray emission appears to be produced by self-Comptonization
(SSC) of the mm/sub-mm synchrotron photons emitted in this region. In this paper, 
we show that the prominent X-ray flare seen in Sgr A* may be due to a sudden 
enhancement of accretion through the circularized flow. Depending on whether 
the associated response of the anomalous viscosity is to increase or decrease 
in tandem with this additional injection of mass, the X-ray photons during the 
outburst may be produced either via thermal bremsstrahlung (if the viscosity 
decreases), or via SSC (if the viscosity increases).  However, the latter 
predicts a softer X-ray spectrum than was seen by {\it Chandra}, so it appears 
that a bremsstrahlung origin for the X-ray outburst is favored. A strong 
correlation is expected between the mm/sub-mm and X-ray fluxes when the flare 
X-rays are produced by SSC, while the correlated variability is strongest 
between the sub-mm/far-IR and X-rays when bremsstrahlung emission is dominant 
during the flare. In addition, we show that future coordinated multi-wavelength 
observations planned for the 2002 and 2003 cycles may be able to distinguish 
between the accretion and jet scenarios.

\end{abstract}




%

\section{Introduction}

The unusual radio source Sgr A* appears to be the radiative manifestation
of the ``dark'' matter concentration at the Galactic center (e.g., Melia
\& Falcke 2001). Discovered by Balick \& Brown in 1974, this object is a 
bright compact source of cm to mm/sub-mm waves, and appears to anchor the 
stars swarming around it with velocities in excess of one thousand kilometers
per second on orbits with a radius no bigger than about ten light days 
(Ghez et al. 2000).  Despite this compelling evidence that a supermassive
black hole accounts for most, if not all, of the inferred $2.6\times 10^6\;
M_\odot$ at the Galactic center, the orbits traced by the stars near 
Sgr A* are approximately $30,000$ times larger than the predicted size of
its event horizon, so alternative explanations cannot yet be ruled out on 
this basis alone.  However, the situation has changed dramatically with the 
recent {\it Chandra} detection of an X-ray flare from this source (Baganoff 
et al. 2001). The ten-minute variability seen during the three-hour event 
argues for an emitting region no bigger than about 20 Schwarzschild radii
(based on light-travel limitations), 
constraining the volume within which the matter is compressed by a factor 
1,500 better than previous studies (Melia 2001).  (For Sgr A*, the 
Schwarzschild radius $r_S\equiv 2GM/c^2$ is approximately $7.7\times 10^{11}$ 
cm, or about $1/20$ A.U. The gravitational radius $r_g\equiv r_S/2$ will
be used to scale the relevant physical quantities throughout this paper.)

As we shall see below, the characteristics associated with this flare appear 
to be consistent with a picture in which the dominant emission region, at least 
at mm/sub-mm and X-ray wavelengths, is associated with a compact Keplerian flow 
of hot, magnetized gas within $\sim 12$ gravitational radii of the black hole
(Melia et al. 2000; Bromley et al. 2001).  This structure may be the inner 
portion of what forms after the low angular momentum gas (with specific angular 
momentum $l\sim 60\,c\,r_g$) captured by Sgr A* circularizes at $\sim 60-100\,r_g$ 
(see, e.g., Coker \& Melia 1997). Earlier, we showed that the inner $\sim 12\ r_g$ 
of such a configuration could not only produce the mm to sub-mm bump in Sgr A*'s 
spectrum (Falcke et al. 1998; Melia et al. 2001), but that it could also account for its 
linear polarization properties (Aitken et al. 2000; Bower et al. 1999; Melia et al. 2000;
Bromley et al. 2001).  Sgr A*'s cm radio emission may instead be due to non-thermal
synchrotron processes in the circularization zone, where the flow evolves from
quasi-spherical accretion at large radii to a Keplerian structure further in.
The {\it quiescent}-state X-ray emission detected with {\it Chandra} appears to 
be produced via synchrotron self-Comptonization (SSC) of the mm/sub-mm photons (Liu \& 
Melia 2001). We note, in this regard, that the latest high-energy observation of Sgr A*
appears to be at odds with the much flatter spectrum predicted by large, two-temperature
ADAF disks (Narayan et al. 1995).

The flare lasted about three hours; during that time, the hard band X-ray flux
decreased by a factor of five in less than ten minutes, consistent with the 
viscous time scale for the inner $6\ r_g$ of the infalling gas, where most of
the high-energy radiation is produced. The dynamical 
time scale associated with this region is typically shorter, 
therefore suggesting that the variability during the outburst might be due to 
an accretion instability within the circularized flow.  This points to a highly 
dynamic event, for which a detailed numerical simulation is required for proper 
modeling.  Nonetheless, the fact that the relevant time scale for establishing 
equilibrium (i.e., the viscous time scale) in the inner few gravitational radii
is much shorter than the duration of 
the flare means that we can effectively adopt quasi-steady conditions during 
the state of enhanced accretion. The initial modeling of such an event is 
therefore straightforward since the structure of the emitting region depends 
primarily on this mass accretion rate $\dot M$. In this {\it Letter}, we explore 
a range of possible system configurations that could account for the observed 
outburst. We demonstrate that the flare can be produced either via thermal 
bremsstrahlung if the anomalous viscosity decreases with the enhanced $\dot M$, 
or via synchrotron self-Comptonization if the anomalous viscosity increases. 
These scenarios make distinct predictions concerning Sgr A*'s flare spectrum 
and the flux density correlations at different wavelengths, which may be 
tested with future coordinated broadband observations. 

\section{Physics of the Transient Event}

Several characteristics associated with the X-ray flare stand out, setting very 
strict constraints on the nature of this event (Baganoff et al. 2001). First, the 
flare lasted about three hours and, near its middle, the $4.5-8$ keV luminosity 
dropped abruptly by a factor of 5 in 10 minutes.  The $2-4.5$ keV luminosity followed 
a similar pattern, though its drop was less sharp, and appeared to lag that of the 
hard X-rays by a few minutes.  Second, the peak of the flaring-state had a luminosity 
45 times greater than that of the quiescent-state.  This huge enhancement in power
suggests that a severe change occurred in the physical properties of the emitting gas,
effectively involving the {\it whole} region. Third, the flare-state spectrum had a spectral 
index of $0.3^{+0.5}_{-0.6}$, which is much harder than that in the quiescent-state 
(i.e., $1.2^{+0.5}_{-0.7}$). This contrasts with the prediction of the disc-corona model 
for AGNs (see, e.g., Ulrich et al. 1997). All these features are quite unique to the 
X-ray flare in Sgr A*.  

The anomalous viscosity is given as $\nu\equiv{(2/3)}{W_{r\phi}/\Sigma\;\Omega}$,
where $\Sigma$ is the column density, $\Omega$ is the angular velocity, and
$W_{r\phi}$ is the vertically integrated sum of the Maxwell and Reynolds 
stresses (Balbus et al. 1994).  For the problem at hand, the Maxwell stress 
dominates, and $W_{r\phi}\approx\beta_\nu\int dz\;\langle{\beta_p\,P}\rangle$
is adequately described by two magnetic parameters in this
model: $\beta_\nu$ is the ratio of the stress to the magnetic field energy density,
and $\beta_p$ is the ratio of magnetic energy density to thermal pressure $P$. The 
radial velocity is given by $v_r\sim (4\beta_p\beta_\nu/9)\,(GM/r)^{1/2}$,
assuming the gas temperature attains its virial value. From this, 
we get the viscous time scale for the gas in the inner few gravitational radii of 
the disk: $\tau_v\equiv r_g/v_r\sim 9.6\,(r/r_g)^{1/2}(0.05/\beta_p\beta_\nu)$ mins. 
We note that $\tau_v$ is consistent with the variability time scale ($\sim
10$ mins) within the flare when $\beta_p\beta_\nu\sim 0.05$, which is close to 
the results produced in detailed MHD simulations (Brandenburg et al. 1995; 
Hawley et al. 1996), when one takes into account the various approximations
made in these calculations, their limited spatial resolution and the differences 
of the physical conditions between their simulations and the accretion model
adopted here. 

Under the same conditions, the dynamical time scale for gas flowing within 
the inner Keplerian region is $\tau_d=2\pi r/v_k$, where $v_k=(GM/r)^{1/2}$ 
is the azimuthal velocity at radius $r$. Its scaled value $\tau_d\approx 
1.3\,(r/r_g)^{3/2}$ mins suggests that a non-equilibrium process may be
responsible for initiating the injection or depletion of matter through
the inner orbits, which then results in an overall fluctuation on a viscous 
time scale as the disk readjusts.  As is well known by now (see, e.g.,
Melia et al. 1992), most of the flux from Sgr A* at a given frequency is
produced by gas in a relatively narrow range of radii (with the highest energy 
radiation being produced in the most compact regions).  Thus, whereas the
duration of a fluctuation probably corresponds to the time required for viscosity 
to re-establish equilibrium, the overall extent of the flare may have been 
attributable to certain characteristics of the infalling plasma (perhaps its 
spatial extent, or its clumping profile). There are several possible 
non-equilibrium processes that could have started the dip near the middle of 
the burst, including a dynamo-induced magnetization of the orbiting plasma.  
This can happen over one orbital period, i.e., $\tau_d$ (see, e.g., Balbus 
et al. 1994; Melia, Liu, \& Coker 2001).  This could, among other things, 
result in a rapid, though transient, increase in the anomalous viscosity, 
followed by a rapid draining of the inner disk. Other mechanisms include 
the intriguing possibility that the infalling plasma may by comprised of 
clumps with a variety of specific angular momenta, so that the material falling in 
at later times effectively cancels (or reverses) the angular momentum of the 
material already in orbit about the black hole. This too can lead to a temporary
thinning of the inner disk.  Unfortunately the photon statistics during the 
flare were not of sufficient quality for the spectrum to be determined as a 
function of time;  only data associated with the integrated flux are available,
so spectral information that can help us to discern between these effects 
near the middle of a burst must await future observations.

With this assessment, we conclude that a $\sim 12\,r_g$ hot, magnetized 
Keplerian flow can account for the temporal behavior of the X-ray flare 
quite naturally.  Adopting quasi-steady conditions in the accretion model 
for Sgr A* (Melia et al 2001), we note that the structure of the disk is then 
determined primarily by $\dot M$ and the anomalous viscosity through this region. 
The inner boundary condition is chosen such that the stress is zero there. At
the outer boundary $r_o$, we also need to know the gas temperature $T_o$. But 
this is not difficult to constrain in cases where the inflowing gas is emitting 
inefficiently before it circularizes (which appears to be the case for Sgr A*); 
we would expect that $T_o$ should be close to its virial value at that radius, 
i.e., $T_o\sim 2GM\,m_p/9kr$, where $m_p$ is the proton mass and $k$ is the 
Boltzmann constant.

Figure 1 shows the profiles of temperature and density as functions of radius
for the inner $\sim 12\,r_g$ of the Keplerian region in the quiescent state, based
on the hot, magnetized disk model described above.  The spectral fit corresponding
to these conditions is represented by the thin solid curve in Figure 2 (see also
Liu \& Melia 2001).  For this, and all other, models discussed in this paper, 
the Keplerian structure has an inclination angle of $45^\circ$ to the line
of sight, an inner boundary $r_i$ 
of $2.4\ r_g$ (we use Newtonian geometry for these calculations---the full
relativistic treatment will be incorporated into the more detailed numerical
simulations to follow) and an outer boundary of $12\ r_g$. The outer boundary 
temperature is set equal to 45 percent of its virial value. The best fit model 
for the quiescent emission has an accretion rate of $7\times 10^{16}$ g
s$^{-1}$ and a viscosity parameter $\beta_\nu =1.0$.  For a plasma with
these characteristics, there are essentially three dominant radiation mechanisms:
thermal synchrotron produces the mm to sub-mm bump; self-Comptonization of these
low-energy photons accounts for the quiescent-state X-ray spectrum; finally,
bremsstrahlung can be significant under some conditions.  For this particular
fit, the high temperature and low number density result in a thermal 
bremsstrahlung flux density smaller than $10^{-10}$ Jy at all frequencies,
rendering its contribution negligibly small compared to SSC.

The success of this model in accounting for Sgr A*'s linear polarimetric characteristics 
(see, e.g., Bromley et al. 2001), together with the natural association we can
make between the viscous time scale and the variability of the observed flare, 
supports the view that the outburst could well have been produced by a transient 
enhancement of $\dot M$ through the inner Keplerian region.  According to this
hot, magnetized accretion model, several important changes are expected to 
result from the injection of new matter into the system.  First, the increase 
in $\dot M$ will increase the particle number density in the flow, which implies 
more efficient cooling and thus a lower temperature. The scale height of the 
plasma will decrease accordingly, which makes the number density even bigger. 
Consequently, the radio emission will become optically thin at a higher
frequency. (Note, however, that an enhanced accretion rate may also affect the
inner boundary condition, which we ignore in this first pass through the
problem. If an additional detailed exploration of this picture is warranted by
future observations of Sgr A*, this effect will be included, along with an 
incorporation of the general relativistic corrections.)
Second, the anomalous viscosity is also expected to change, though
it is not yet understood whether $\beta_p$ and $\beta_\nu$ will increase
or decrease.  In this paper, we therefore consider both circumstances to bracket
the range of possible outcomes. For example, Figure 1 shows what happens if
$\dot M$ is increased, corresponding to two rather diverse responses of the 
anomalous viscosity.  The dotted curves represent the physical variables
when the viscosity parameter $\beta_\nu$ decreases by a factor of about 
two to $0.481$. All the other parameters are the same as in the quiescent
state.  The decrease in the anomalous viscosity increases the number density even
further, and the inflowing gas cools down to about $10^9$ K at small radii. 
The decrease in $n_B$ as the gas approaches the inner boundary is due to the 
zero stress condition there. The dashed curves demonstrate the behavior of $T$ and
$n$ when $\beta_p$ increases by a factor of about seven compared to its value in
the quiescent state.  With an increase in the anomalous viscosity, the effects
of an enhanced $\dot M$ on $n_C$ are mostly annulled, and the structure of 
the inflow is therefore similar to that of the pre-flare state. 

\section{Results and Discussion}

The best fit model for an accretion-induced flare in Sgr A* is shown in Figure 2,
which also shows the quiescent-state spectrum for comparison. In this case,
the enhancement in $\dot M$ is coupled with an associated decrease in anomalous 
viscosity, the combination of which results in an increase of the particle
number density by a factor of 30 at the outer boundary.  The structure of
the Keplerian flow consequently changes considerably (Fig. 1). All three 
of the emission processes contribute significantly to the overall spectrum. 
An associated strong sub-mm/far-IR flare is expected during the X-ray outburst.
Note, however, that due to a sharp decrease in the gas temperature, the mm 
flux is not expected to change significantly; depending on the exact model
parameters, it may even decrease slightly.  It is not yet certain, though,
whether this can lead to an anti-correlation between the mm and X-ray
flux densities, since the emission from the circularization zone further out
may also contribute to the mm spectrum (see also Melia 1992, 1994). 
Unfortunately, the physics of this region is not yet well understood. Future 
high-resolution numerical simulations of the time-dependent problem may resolve 
this ambiguity. For this situation, the high-energy X-ray emission is dominated 
by bremsstrahlung processes and the spectrum flattens during the flare. 
We note, however, that SSC may still contribute some flux to the soft X-ray band.
This is interesting in view of the fact that a power-law fit to the flare-state
spectrum suggests less absorption than expected, implying a low column density.
This excess soft X-ray emission may be due to the additional flux produced by 
SSC above pure bremsstrahlung up to $\sim 4$ keV. 
In this paper, we have adopted a quasi-equilibrium assumption for the 
flare state.  The flare's actual time profile may be reproduced by solving 
the full time-dependent dynamical equations describing the accretion flow, 
taking into account the possible asymmetry and diffusion induced by the 
additional mass injection. 

Synchrotron self-Comptonization can also produce significant X-ray emission
under some circumstances, specifically, when an increase in $\dot M$ is associated
with an enhanced anomalous viscosity. Due to a decrease in gas temperature, however, 
the peak of the Compton-scattered spectrum is not expected to shift significantly
toward higher frequencies compared to the quiescent state. So the X-ray spectral 
index should change only marginally from its value in the quiescent state.  This 
does not appear to be consistent with what has been observed.  These features are 
evident in Figure 3, which shows the best fit spectrum when SSC dominates the X-ray 
flux during the flare; the predicted X-ray spectrum is simply too soft. 

This situation is quite different from that of the jet model, in which 
SSC is invoked as the sole mechanism
for producing the X-ray flare (Markoff et al. 2001). In this alternative picture, 
the gas temperature is a free parameter, and is arbitrarily increased by a factor 
of three or four to fit the flare-state spectrum. The increase in $T$ has the
dual effect of shifting the peak of the Comptonized component by more than an
order of magnitude to higher frequencies, and flattening and extending the mm
spectrum. The combination of these effects makes it possible for the peak of
the Comptonized emission to be shifted into the X-ray domain, thereby accounting
for a flattening of the spectral index during the flare.  In the accretion picture, 
the enhanced $\dot M$ suppresses the temperature sufficiently for the Comptonized 
component to peak at UV energies.  Aside from these differences, the jet and 
accretion models also differ in their predictions for the associated flare at 
other wavelengths.  The accretion model suggests that there should be a correlated 
strong sub-mm/far-IR flare, whereas the jet scenario predicts a strong IR burst.  
Coordinated observations at mm/sub-mm and X-ray
energies and at X-ray and $\gamma$-ray energies are now being planned for
the 2002 and 2003 observing cycles, so these differences may soon lead to a 
possible resolution of which picture accounts for most of Sgr A*'s radiative
emission. 

{\bf Acknowledgments} 
We are grateful to Fred Baganoff, Marco Fatuzzo and Mark Morris for very 
valuable discussions.  The anonymous referee improved the manuscript
significantly with his thoughtful comments. This research was partially 
supported by NASA under grants NAG5-8239 and NAG5-9205, and has made use 
of NASA's Astrophysics Data System Abstract Service.  FM is very grateful 
to the University of Melbourne for its support (through a Miegunyah Fellowship). 

{}

%
%

\begin{figure}[thb]\label{fig1.ps}
{\begin{turn}{0}
\epsscale{0.8}
\centerline{\plotone{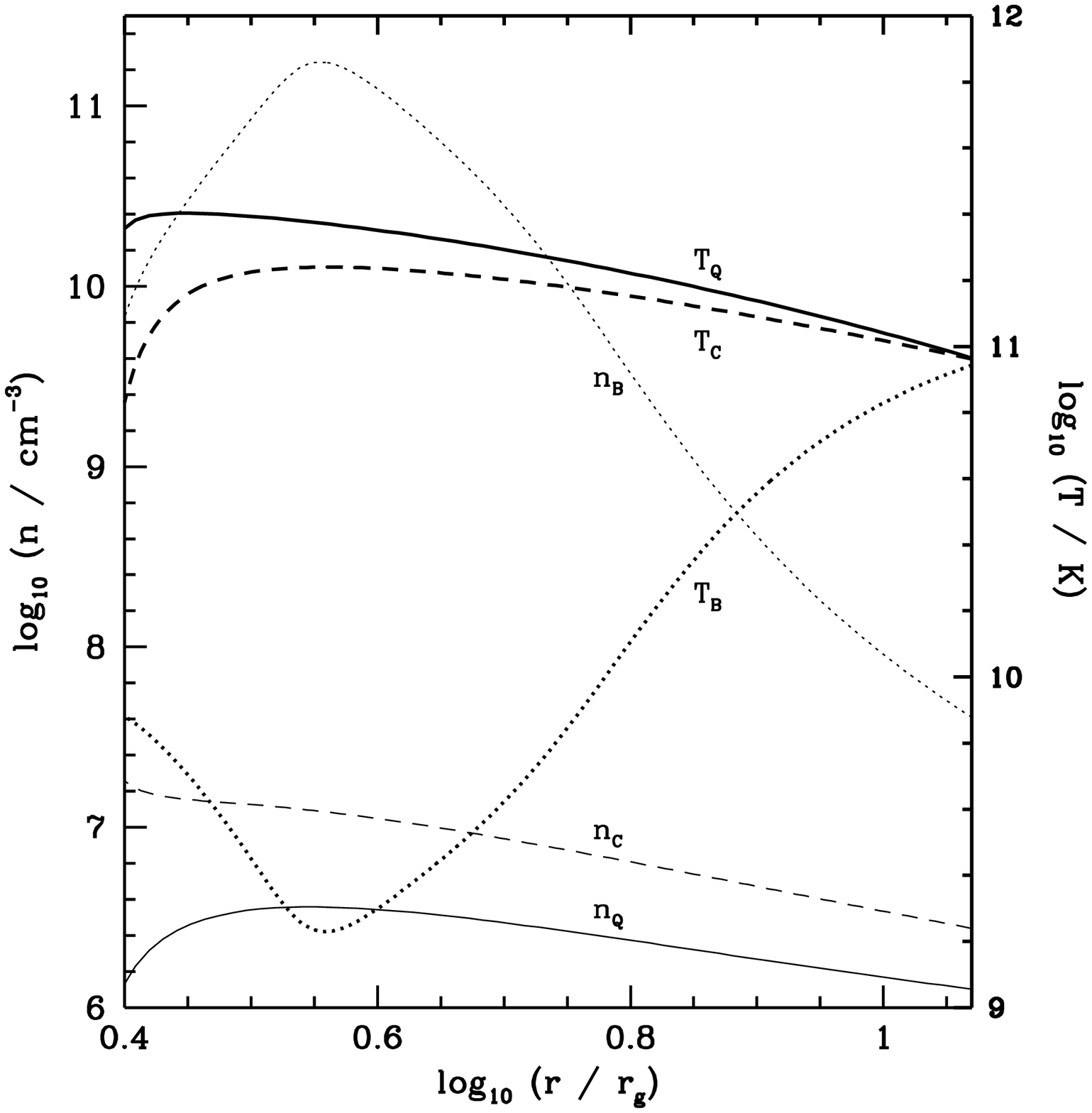}}
\end{turn}}
\caption{Radial profiles of the temperature (thick curves) and the number density
(thin curves) of the Keplerian gas for three sets of physical parameters. In all
cases, the ratio of the outer boundary temperature to its virial value is 0.45 
and an inclination angle of $45^\circ$ is assumed for the calculation of the
spectrum. The solid curves show the best fit model for the quiescent state
of Sgr A* (with temperature $T_Q$ and number density $n_Q$). 
The accretion rate in this situation is $7\times 10^{16}$ g
s$^{-1}$ and $\beta_\nu=1.0$. The other parameters are $\beta_p=0.09$,
$r_o=12r_g$, and $r_i=2.4r_g$. In Fig. 2, we will present the best fit model
for the flare state, which corresponds to the dotted curves shown here
(labeled $T_B$ and $n_B$). The fit shown in Fig. 3, where the X-rays 
are produced via synchrotron self-Comptonization, correspond to the 
dashed curves used here (labeled $T_C$ and $n_C$).} 
\end{figure}

\begin{figure}[thb]\label{fig2.ps}  
{\begin{turn}{0}
\epsscale{0.8}
\centerline{\plotone{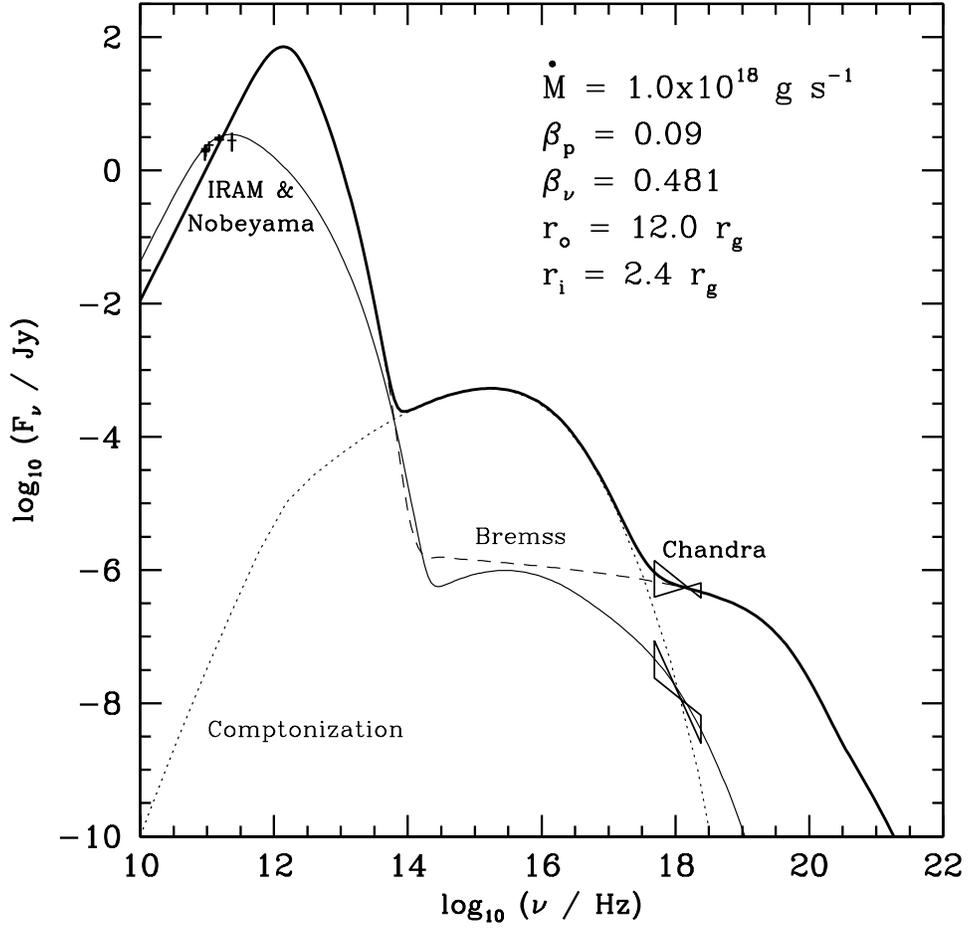}}
\end{turn}}
\caption{Best fit spectrum (thick solid curve) for the flare state, produced 
primarily with thermal bremsstrahlung emission. The physical parameters are 
listed in the figure.  The thin solid curve corresponds to the best fit for
the quiescent-state spectrum, corresponding to the dotted curves used in Fig. 1.}
\end{figure}

\begin{figure}[thb]\label{fig3.ps}
{\begin{turn}{0}
\epsscale{0.8}
\centerline{\plotone{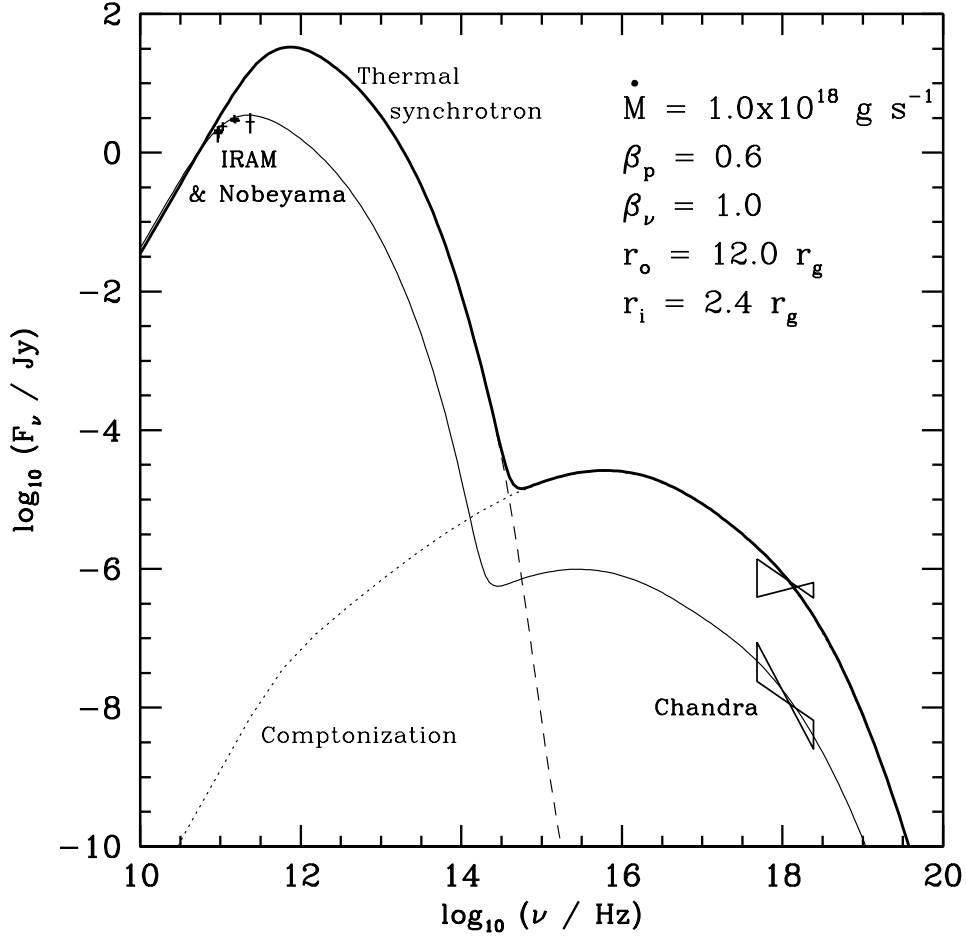}}
\end{turn}}
\caption{A fit to the flare-state spectrum when synchrotron self-Comptonization
dominates (thick solid curve). The physical parameters are quoted in the figure. 
The thin solid curve corresponds to the best fit for the quiescent-state spectrum,
corresponding to the dashed curves used in Fig. 1.
In this case, the predicted X-ray flare spectrum is not a good match to the
{\it Chandra} data.}
\end{figure}

\end{document}